%Paper: gr-qc/9409050
%From: mannheim@main.phys.uconn.edu (Philip Mannheim)
%Date: Fri, 23 Sep 94 11:21:02 EDT

\magnification=1000
\baselineskip=0.54truecm
\centerline{\bf HIGGS MECHANISM AND THE STRUCTURE OF THE ENERGY-MOMENTUM
TENSOR}
\centerline{\bf IN EINSTEIN GRAVITY AND CONFORMAL GRAVITY}
\vskip
0.30truecm
\centerline{\bf Philip D.
Mannheim}
\vskip
0.10truecm
\centerline {Department of Physics, University of Connecticut, Storrs,
CT 06269-3046}
\vskip 0.05truecm
\centerline{(mannheim@uconnvm.uconn.edu)}
\vskip
0.05truecm
\centerline{and}
\vskip 0.05truecm
\centerline{\bf Demosthenes Kazanas}
\vskip
0.10truecm
\centerline{Laboratory for High Energy Astrophysics,
NASA/Goddard Space Flight Center, Greenbelt, MD 20771}
\vskip 0.05truecm
\centerline{(kazanas\%lheavx.dnet.nasa.gov)}
\vskip 0.10truecm
 \centerline{preprint UCONN-94-7, September 1994, to appear in the
 proceedings of the 7th Marcel Grossmann Meeting}
\vskip 0.10truecm
\baselineskip=0.44truecm
\noindent
{\bf Abstract}$~~~~$In the standard treatment of the Einstein
gravitational theory the energy-momentum tensor has always been taken to be
composed of
perfect fluid aggregates of kinematic Newtonian point test particles with
fundamental mechanical masses. Moreover, this standard prescription was not
revised after the
discovery of the mass-generating Higgs mechanism which is known to be present
in the elementary
particle  physics of these self-same sources and which is also required in the
conformal invariant
gravitational alternative being considered by Mannheim and Kazanas.
In this short contribution we show that despite the presence of the Higgs
mechanism, the standard
geodesic motion and Euler hydrodynamics still obtain in the one-particle sector
of the
theory even
while the overall energy-momentum tensor differs substantially from the
conventional kinematic one.
\vskip 0.10truecm
\baselineskip=0.54truecm

During the last few years Mannheim and Kazanas$^{1-9}$
have undertaken a study of a completely conformal invariant four-dimensional
theory of gravity, based on invariance of the geometry under stretchings
$g_{\mu \nu} (x) \rightarrow \Omega^{2}(x) g_{\mu\nu}(x)$
and associated unique gravitational action $I_W =-2\alpha \int
d^4x(-g)^{1/2}(R_{\lambda\mu}R^{\lambda
\mu}-(R^{\alpha}_{\phantom{\alpha} \alpha})^2 /3)$
with purely dimensionless coupling constant $\alpha$, and
considered it as a fully relativistic metric based candidate alternative
to the standard Newton-Einstein theory. This study has yielded the
exact classical interior and exterior solutions associated with a static
spherically symmetric gravitational source in the theory. One of the
general features of these solutions is that besides recovering the Newtonian
$-\beta/r$
potential, they also yield a second potential term $\gamma r/2$ which grows
linearly with distance. Using this $V(r)=-\beta/r+\gamma r/2$ form for the
potential, conformal gravity is then able$^{7}$ to fit the observed
galactic rotation curves of some representative spiral galaxies by integrating
this potential over their observed surface luminosities.
The fits obtained to these rotation curves  rival those made under the
assumption of dark matter halos and point to
alternative gravity as a possible explanation for the observed
systematics of galactic rotation curves.

While the fitting to the rotation curves is encouraging, we note that once we
invoke conformal invariance, we do not merely change the gravitational action
(and
thus the gravitational equations of motion) but also we radically modify the
structure of the energy-momentum tensor which is to serve as its source, with
sources no longer being describable as test particles or as their perfect fluid
aggregates (i.e. 'billiard balls'). Rather, the sources are now matter fields
(whose particle content only emerges after second quantization)
which must be accompanied by Higgs fields if the matter fields are to acquire
masses in an otherwise scale invariant theory, with the energy and momentum of
these Higgs fields then providing an explicit additional contribution to the
gravitational
source which cannot be ignored. (In the standard model of course it is ignored
simply because it is 120 orders of magnitude too big). Thus
the radical departure of dynamical mass generation (in any theory in fact) is
that it automatically
excludes the  whole standard Newtonian picture of mechanical point test
particles and perfect fluid
sources as well as its associated intuition. Moreover, even though relativists
generally
choose not to do so, in the standard Einstein theory itself we are nonetheless
still obliged to
discard this Newtonian prescription for gravitational sources once mass is in
fact generated by
the Higgs  mechanism. However, since geodesic motion and the standard Euler
hydrodynamic theory enjoy considerable experimental support (this presumably
being the reason
why Newtonian sources are still in favor),
it becomes necessary to recover these features
in any theory where there is in fact
dynamical mass generation. This we do in this contribution, and as we shall
see, our	study will expose both the level of validity and the shortcomings of
the standard Newtonian picture of sources.

To formulate the problem we consider the most general matter
action for generic scalars $S(x)$ and fermions
$\psi(x)$ coupled conformally to gravity, viz.
$$I_M=-\int
d^4x(-g)^{1/2}[{1 \over 2}S^\mu S_\mu+\lambda S^4-{1 \over
12}S^2R^\mu_{\phantom
{\mu}\mu}+i\bar{\psi}\gamma^{\mu}(x)(\partial_\mu+\Gamma_\mu(x)) \psi
-hS\bar{\psi}\psi] \eqno(1)$$
\noindent
where $\Gamma_{\mu}(x)$ is the fermion spin connection and $h$ and
$\lambda$ are dimensionless couplings constants. (In most treatments of the
Higgs
mechanism in particle physics the Higgs field is minimally
coupled to matter so that the $-S^2R^\mu_{\phantom{\mu}\mu}/12$ term is absent.
The
conclusions we reach below are insensitive to this term, so we carry it
here to retain full
conformal symmetry). For our action the
energy-momentum tensor takes the form
$$T_{\mu \nu} = i \bar{\psi} \gamma_{\mu}(x)[ \partial_{\nu}
+\Gamma_\nu(x)]
\psi+{2 \over 3}S_\mu S_\nu  -{1 \over 6}g_{\mu\nu}S^\alpha S_\alpha
-{1 \over 3}SS_{\mu;\nu}
+{1 \over
3}g_{\mu\nu}SS^\alpha_{\phantom{\alpha};\alpha}
-{1 \over
6}S^2(R_{\mu\nu}-{1 \over
2}g_{\mu\nu}R^\alpha_{\phantom{\alpha}\alpha})
-g_{\mu\nu}\lambda S^4
\eqno(2)$$
\noindent
when the generic matter fields obey the equations of motion
$$S^\mu
_{\phantom{\mu};\mu}+{1 \over 6}SR^\mu_{\phantom{\mu}\mu}
-4\lambda S^3+h\bar{\psi}
\psi=0\eqno(3)$$
$$
i\gamma^{\mu}(x)[\partial_{\mu}
+\Gamma_\mu(x)]
\psi - h S \psi =
0.\eqno(4)$$
\noindent
With use of Eqs. (3) and (4) one can show that $T^{\mu \nu}$ is both
covariantly conserved and covariantly traceless as is to be expected in a
conformal invariant theory, with this conformal $T^{\mu \nu}$ possessing a
structure radically different from that associated with a mechanical
Newtonian test particle. As we can also see, the Ricci scalar term in the
scalar field equation of motion can serve to provide a minimum to the scalar
field potential away from the origin (even without any fundamental
non-conformal invariant tachyonic
$-\mu^2S^2$ Higgs mass term) at some value $S_0$ which we take (after a
conformal transformation $S(x)\rightarrow \Omega^{-1}(x) S(x)$
as necessary) to be a spacetime
constant. Thus curvature itself can serve to spontaneously break the
conformal symmetry, to thus enable the fermion to acquire an induced
dynamical mass $m=hS_0$, and to suggest that the standard Higgs mechanism may
have a geometrical basis.

When the scalar field is taken to be a constant the matter
field energy-momentum tensor simplifies to
$$T_{\mu \nu} = i \bar{\psi} \gamma_{\mu}(x)[ \partial_{\nu}
+\Gamma_\nu(x)]\psi
-
{1 \over 6}S_0^2(R_{\mu\nu}-{1 \over
2}g_{\mu\nu}R^\alpha_{\phantom{\alpha}\alpha})
-g_{\mu\nu}\lambda S_0^4 =T^{kin}_{\mu \nu}+T_{\mu \nu}(S_0)
\eqno(5)$$
\noindent
using an obvious notation. Now while it was already noted that the full
$T^{\mu\nu}$ is covariantly conserved, we can now see that two separate
pieces of it are also independently conserved, namely $T^{\mu \nu}(S_0)$
(immediately because of the Bianchi identity which the regular Einstein
tensor obeys) and thus consequently also $T^{kin}_{\mu \nu}$ (as could also
be shown directly from the Dirac equation of Eq. (4)). Hence,
and this is the key point, while the
purely kinematic fermionic $T^{kin}_{\mu \nu}$ could in principle share
energy and momentum with the Higgs field that gives it its mass (the
situation one would initially have to expect given only the conservation of
the full $T^{\mu \nu}$), in fact we see that it does not. Moreover, this
fact is made all the more remarkable when one realizes that neither of these
two pieces is separately covariantly traceless, but only their sum, so
that the fermion still needs its Higgs field to maintain the
tracelessness of the overall energy-momentum tensor, even as it does
not share energy and momentum with it.

Now as far as the pure fermionic sector of the theory is concerned, we note
that
its relevant equations, namely the Dirac equation of Eq. (4) with $S(x)=
S_0$ and the covariant conservation condition on $T^{kin}_{\mu \nu}$, are
exactly
the same as those one would have written down if one had simply started with
a gravitational source consisting of a purely kinematic fermion field with a
fundamental non-conformal invariant mechanical mass $m$ instead. Hence the one
particle excitations of the fermion in a spontaneously broken Higgs theory
are identically the same as they would have been if the fermion simply had a
mechanical bare mass; with the associated covariant conservation of
$T_{\mu\nu}^{kin}$ then giving geodesic motion for such particles in an
external
gravitational field just as it does in the standard theory where $T^{kin}_{\mu
\nu}$
is ordinarily taken to be the entire gravitational source. Further, an
incoherent averaging over a bath of the normal mode solutions to Eq. (4) then
gives the fermion kinetic term the form of a kinematic perfect fluid, viz.
$T^{kin}_{\mu\nu}=(\rho+p)U_\mu U_\nu+pg_{\mu\nu}$ where
$\rho= \int dk k^2 E_k/ \pi^2$,  $p=\int dk k^4/ 3 \pi^2 E_k$, with its
covariant
conservation then giving$^{6,8}$ the familiar Euler hydrostatic equilibrium
equation when the fermions are the source of a weak gravitational field.
Hence we see that both geodesic motion and perfect fluid hydrodynamics
still occur in theories with dynamical mass generation, and that in and of
itself,
the behavior of the one-particle sector of the theory is
simply a totally inadequate guide as to the structure of the full
energy-momentum tensor of gravitational theory since the Higgs field
remarkably manages to hide itself from the motions of the one-particle
excitations.
The fermionic one-particle sector of the Higgs theory thus has the same
structure as the one
thought to exist in the pre-Higgs era when the standard Einstein theory was
first written down, a structure which curiously is still thought to prevail
for the entire $T^{\mu \nu}$ in the standard theory despite the subsequent
discovery of dynamical mass generation. However, unfortunately for the standard
picture, even while
the one-particle motions may in fact decouple from the Higgs fields in the
energy and momentum
balance, the gravitational  field itself is still very much sensitive to these
Higgs fields as
gravity couples to the  zero of energy and not merely to the change in energy
detectable in the
one-particle sector. Since this latter Higgs energy and the attendant back
reaction on the
geometry contained in $T^{\mu\nu}(S_0)$ are simply ignored in the standard
Newtonian based
picture of gravitational sources (even though $T^{\alpha}_{\phantom {\alpha}
\alpha}(S_0)$
is equal to $-T^{\alpha}_{\phantom {\alpha}\alpha}(kin)$ and thus just as large
- neither smaller
nor larger, and certainly not 120 orders of magnitude larger, thereby enabling
an underlying
conformal structure to resolve the cosmological constant problem), at  the
present time the
entire standard treatment of sources in the Einstein theory  must be regarded
as suspect; and a
reader who may wish to reject the implications of conformal invariance as far
as the gravitational
sector is concerned, is still not free to ignore the back reaction of the Higgs
fields on the
geometry when considering the standard Einstein theory itself.
\smallskip
%\noindent {\bf References}
%\smallskip
 \baselineskip=0.49truecm
\item{1.}
         P. D. Mannheim, {\it Gen. Rel. Grav.} {\bf  22}, 289 (1990).

\item{2.}
         P. D. Mannheim and D. Kazanas, {\it Ap. J.} {\bf 342}, 635 (1989).

\item{3.}
         D. Kazanas and P. D. Mannheim,  {\it Ap. J. Suppl. Ser.} {\bf 76}, 431
(1991).

\item{4.}
         P. D. Mannheim and D. Kazanas, {\it Phys. Rev.} {\bf D44}, 417 (1991).

\item{5.}
         P. D. Mannheim, {\it Ap. J.}  {\bf 391} 429 (1992).

\item{6.}
         P. D. Mannheim, {\it Gen. Rel. Grav.} {\bf 25}, 697 (1993).

\item{7.}
         P. D. Mannheim, {\it Ap. J.}  {\bf 419}, 150 (1993).

\item{8.}
         P. D. Mannheim and D. Kazanas, {\it Gen. Rel. Grav.} {\bf 26}, 337
(1994).

\item{9.}
         P. D. Mannheim,  {\it Founds. Phys.} {\bf 24}, 487 (1994).

\end